\documentclass[twocolumn,showpacs,preprintnumbers,amsmath,amssymb]{revtex4}

\usepackage{graphicx}
\usepackage{dcolumn}
\usepackage{bm}

\begin{document}


\title{Multi-Higgs U(1) Lattice Gauge Theory in Three Dimensions}

\author{Tomoyoshi Ono} \author{Ikuo Ichinose}
 \affiliation{Department of Applied Physics, Graduate School of Engineering, \\
Nagoya Institute of Technology, 
Nagoya, 466-8555 Japan 
}
\author{Tetsuo Matsui}%
\affiliation{%
Department of Physics, Kinki University, 
Higashi-Osaka, 577-8502 Japan
}%

\date{\today}

\begin{abstract}
We study the three-dimensional compact U(1) lattice gauge theory
 with $N$ Higgs fields numerically. This model is relevant to
multi-component superconductors, antiferromagnetic
spin systems in easy plane, inflational cosmology, etc. For $N=2$, 
the system has a second-order phase transition line $\tilde{c}_1(c_2)$
in the $c_2$(gauge coupling)$-c_1$(Higgs coupling) plane, 
which separates the confinement phase and the Higgs phase. 
For $N=3$, the critical line is separated into two parts;
one for $c_2 \alt 2.25$ with first-order 
transitions, and the other for
$c_2 \agt 2.25$ with second-order transitions. 
   
\end{abstract}
\pacs{11.15.Ha, 05.70.Fh, 74.20.-z, 71.27.+a, 98.80.Cq}
\maketitle

There are many interesting physical systems involving multi-component
($N$-component) matter fields. 
Sometimes they are associated
with exact or approximate symmetries like ``flavor" symmetry.  
In some cases, the large-$N$ analysis\cite{largen} is applicable 
and it gives us useful information. 
But the properties of the large-$N$ systems may differ
from those at medium values of $N$ that one actually wants to know. 
Study of the $N$-dependence of various systems is certainly interesting 
but not examined well. 

Among these ``flavor" physics, the effect of matter 
fields upon gauge dynamics is of quite general
interest in quantum chromodynamics, strongly correlated 
electron systems, quantum spins, etc.\cite{gauge} 
In this letter,  
we shall study the three-dimensional (3D) U(1) gauge theory
with multi-component Higgs fields $\phi_a(x)
\equiv |\phi_a(x)| \exp(i\varphi_a(x))\
 (a=1,\cdots,N)$.
This model is of general interest, and knowledge of 
its phase structure, order of its phase 
transitions, etc. may be useful to get better 
understanding of various physical systems.
These systems include the following:

{\it $N$-component superconductor:}
Babaev\cite{babaev} argued  that under a high pressure and 
at low temperatures hydrogen gas may become a liquid and
exhibits a transition from a superfluid to a superconductor.
There are two order parameters; $\phi_e$ for electron pairs
and $\phi_p$ for proton pairs. They may be treated as 
two complex Higgs fields ($N=2$). In the superconducting phase, 
both $\phi_e$ and $\phi_p$ develop an off-diagonal long-range 
order, while in the superfluid phase, only the 
neutral order survives; $\lim_{|x| \rightarrow \infty} \langle
\phi_e(x) \phi_p(0)\rangle \neq 0$. 
 
{\it $p$-wave 
superconductivity of cold Fermi gas:}
Each fermion pair in a $p$-wave superconductor
has angular momentum $J=1$ and the order parameter 
has three components, $J_z= -1,0,1$. They are regarded
as three Higgs fields ($N=3$). As the strength of attractive
force between fermions is increased,  a crossover from
a superconductor of the BCS type  to the type of  
Bose-Einstein condensation is expected to take place\cite{ohashi}.

{\it Phase transition of 2D antiferromagnetic(AF) 
spin models:}
In the $s=1/2$ AF spin models, a phase transition occurs from the Neel 
state to the valence-bond solid state as parameters are varied. 
Senthil et al.\cite{senthil}
argued that the effective theory describing
this transition take a form of  
 U(1) gauge theory of spinon ($CP^1$) field $z_a(x)\
(|z_1|^2+|z_2|^2=1)$.
In the easy-plane limit ($S_z=0$), $|z_1|^2=|z_2|^2=1/2$
and so they are expressed by two Higgs fields as
$z_a= \exp(i\varphi_a)/\sqrt{2}$ ($N=2$)\cite{cpn-1}.

Effects of doped fermionic holes (holons) 
to this AF spins are also studied extensively. 
The effective theory obtained by integrating out holon 
variables may be a U(1) gauge theory with $N=2$ Higgs fields
(with nonlocal gauge interactions).
Kaul et al.\cite{kaul} predicts that such a system exhibits
a second-order transition, while numerical 
simulations of Kuklov et al.\cite{kuklov} exhibit 
a weak first-order transition.
This point should be clarified in future study.

{\it Inflational cosmology:}
In the inflational cosmology\cite{guth}, a set of Higgs 
fields is introduced to describe a phase transition
and inflation in early universe.
Plural Higgs fields are necessary
in a realistic model\cite{allahverdi}.

The following simple consideration ``predicts" the phase structure of 
the system.
Among $N$ phases $\varphi_a(x)$ of the Higgs fields,
the sum $\tilde{\varphi}_{+}\equiv\sum_a\varphi_a$ couples
to the gauge field and describes charged excitations, 
whereas the remaining $N-1$ independent linear combinations
$\tilde{\varphi}_{i} (i=1,\cdots,N-1)$ 
describe neutral excitations. The latter $N-1$ modes
may be regarded as a set of $N-1$ $XY$ spin models.
As the $N=1$ compact U(1) Higgs
model stays always in the confinement phase\cite{janke}, 
we expect $N-1$ second-order transitions of the type of the $XY$ model. 

Smiseth et al.\cite{smiseth} studied the {\em noncompact} U(1) Higgs models. 
A duality transformation maps the charged sector into the inverted 
$XY$ spin model. 
Thus they predicted that the system exhibits a single inverted $XY$ transition 
and $N-1$ $XY$ transitions. 
Their numerical study confirmed this prediction for $N=2$.

For $N=2$, Kragset et al.\cite{kragset} studied the effect
of Berry's phase term in the $N=2$ compact Higgs model. 
They reported that Berry's phase term suppresses monopoles
(instantons) and changes the second-order phase transitions
to first-order ones.

In this letter, we shall study the multi-Higgs models
by Monte Carlo simulations. 
We consider the simplest form, i.e., the
3D compact lattice gauge theory without Berry's phase;
the Higgs fields $\phi_{xa}$ are treated in the London limit, 
$|\phi_{xa}|=1$.  
The action $S$ consists of the Higgs coupling
with its coefficient $c_{1a}\; (a=1,\dots, N)$ 
and the plaquette term with its coefficient $c_2$,
\begin{eqnarray}
S&=&{1\over 2}\sum_{x,\mu}\sum_{a=1}^N
\Big(c_{1a}\phi^\dagger_{x+\mu,a}
U_{x\mu}\phi_{xa}+\mbox{H.c.}\Big) \nonumber  \\
&&+{c_2 \over 2}\sum_{x,\mu<\nu}
(U^\dagger_{x\nu}U^\dagger_{x+\nu,\mu}U_{x+\mu,\nu}U_{x\mu}
+\mbox{H.c}),
\label{action}
\end{eqnarray}
where $U_{x\mu}[=\exp(i\theta_{x\mu})]$ is the compact U(1) 
gauge field,
$\mu,\nu(=1,2,3)$ are direction indices (we use them 
also as the unit vectors).

We first study the $N=2$ case with symmetric couplings 
$c_{11}=c_{12} \equiv c_1$.
We measured the internal energy $E\equiv -\langle S \rangle/L^3$
and the specific heat $C\equiv\langle 
(S-\langle S\rangle)^2\rangle/L^3$ in 
order to obtain the phase diagram and determine the order of 
phase transitions,
where $L^3$ is the size of the cubic lattice with the  
periodic boundary condition.

\begin{figure}
\includegraphics[width=8.5cm]{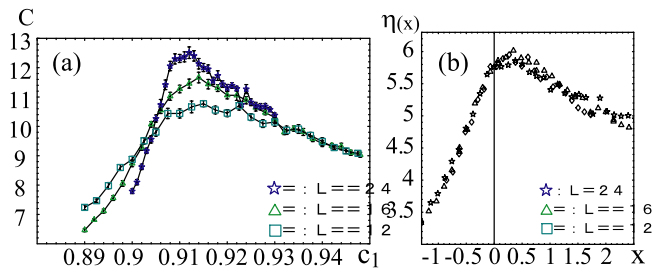}
\caption{\label{fig1}
(a) System-size dependence of specific heat $C$ 
for $N=2$ at $c_2=0.4$.
(b) Scaling function $\eta(x)$ for Fig.\ref{fig1}(a).
}\end{figure}

In Fig.\ref{fig1}(a), we show $C$ at $c_2=0.4$ as a function of $c_1$.
The peak of $C$ develops as the system size is increased. 
The results indicate that a second-order phase transition occurs at 
$c_1 \simeq 0.91$.
By applying the finite-size-scaling (FSS) hypothesis 
to $C$ in the form of 
$C(c_1,L)=L^{\sigma/\nu}\eta(L^{1/\nu}\epsilon)$, where 
$\epsilon=(c_1-c_{1\infty})/c_{1\infty}$ and $c_{1\infty}$ 
is the critical coupling at $L\rightarrow \infty$,
we obtained $\nu=0.67, \; \sigma=0.16$, and $c_{1\infty}=0.909$.
In Fig.\ref{fig1}(b) we plot $\eta(x)$,
which supports the FSS.

The above results for $N=2$ are consistent with the 
``prediction" given above. 
The sum $\tilde{\varphi}_{x+}\equiv \varphi_{x1}+\varphi_{x2}$
couples with the compact gauge field and generates no
phase transition\cite{janke}, while
the difference $\tilde{\varphi}_{x-}\equiv 
\varphi_{x1}-\varphi_{x2}$ 
behaves like the angle variable in the 3D $XY$ model.
The 3D $XY$ model has a second-order phase transition 
with the critical exponent $\nu=0.666...$\cite{XY}. 
Our value of $\nu$ obtained above is very close to this value.
However, it should be remarked that the simple separation 
of variables in terms of $\tilde{\varphi}_{\pm}$ is {\em not}
perfect due to the higher-order terms in the compact gauge 
theory.
Nonetheless, our numerical studies strongly suggests that 
the phase transition for $N=2$ belongs to the universality 
class of the 3D $XY$ model.
 
\begin{figure}
\includegraphics[width=5cm]{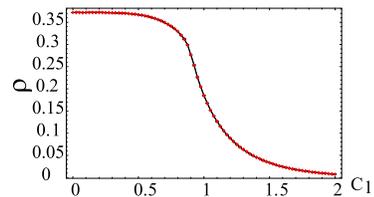}
\vspace{-1.5cm}
\caption{\label{fig2}Instanton density $\rho$ for $N=2$ 
at $c_2=0.4$ as a function of $c_1$. 
}
\end{figure}

It is instructive to see the behavior of the instanton density 
$\rho$. We employ the definition of $\rho$ 
in the 3D U(1) compact lattice gauge theory given by
DeGrand and Toussaint\cite{instanton}.
$\rho$ in Fig.\ref{fig2}
decreases very rapidly near the phase transition 
point.
This indicates that a ``crossover" from dense to dilute instanton 
``phases" occurs simultaneously with the phase transition. 
In other words, the observed phase transition can be interpreted 
as a confinement(small $c_1$)-Higgs(large $c_1$) phase transition.

\begin{figure}
\includegraphics[width=5cm]{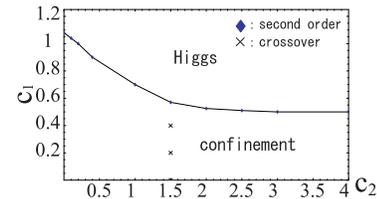}
\caption{\label{fig3}Phase diagram for $N=2$.
There are two phases, confinement and Higgs, separated 
by second-order phase transition line. There also exists 
a crossover line in the confinement phase separating dense 
and dilute instanton-density regions.
}
\end{figure}

In Fig.\ref{fig3}, we present 
the phase diagram for $N=2$ in the $c_2$-$c_1$ plane.
There exists a second-order phase transition line separating 
the confinement and the Higgs phases.
There also exists a crossover line similar to that in the 
3D $N=1$ U(1) Higgs model\cite{janke}.

\begin{figure}
\includegraphics[width=8.5cm]{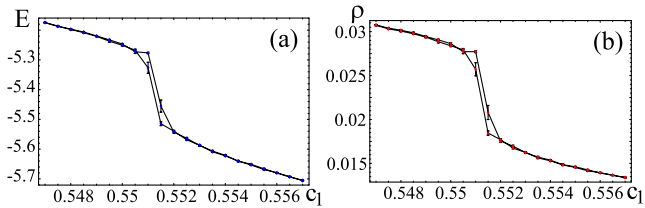}
\caption{\label{fig4}
(a) Internal energy $E$ and (b) instanton density $\rho$ 
for $N=3$ at $c_2=1.5$ and $L=16$.
Both exhibit hysteresis loops indicating a
first-order phase transition at $c_1 \simeq 0.551$.}
\end{figure}

\begin{figure}
\includegraphics[width=8.5cm]{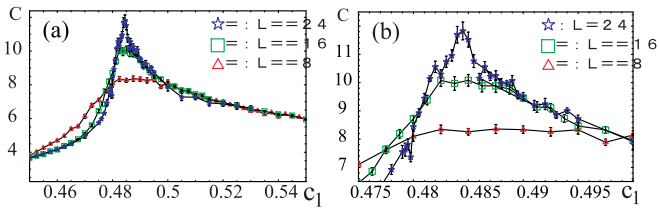}
\caption{\label{fig5}
(a) Specific heat for $N=3$ at $c_2=3.0$. 
(b) Close-up view near the peak. The peak
develops as $L$ increases.}
\end{figure}

\begin{figure}
\includegraphics[width=8.5cm]{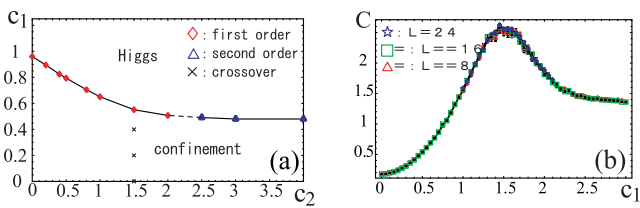}
\caption{\label{fig6}
(a) Phase diagram for the $N=3$ symmetric case.
The phase transitions are first order in the
 region  $c_2 \alt 2.25$, whereas 
they are second order in the region $c_2 \agt 2.25$. 
There exists a 
tricritical point at around $(c_2,c_1)\sim 
(2.25,0.5)$.
Crosses near $c_2 = 1.5$ line show crossovers.
(b) Specific heat for $N=3$ at $c_1=0.2$.
It has a system-size independent smooth peak at which
a crossover takes place.}
\end{figure}

Let us turn to the $N=3$ case. 
Among many possibilities of three $c_{1a}$'s, 
we first consider the symmetric case $c_{11}=c_{12}=c_{13}
\equiv c_1$.
One may expect that there are two ($N-1=2$) second-order 
transitions that may coincide at a certain critical point.
Studying  the $N=3$ case is interesting from a general 
viewpoint of the critical phenomena, i.e., whether 
coincidence of multiple phase transitions changes the 
order of the transition. 
We studied various points in the $c_2-c_1$
plane and found that the order of transition 
changes as $c_2$ varies.
In Fig.\ref{fig4}, we show $E$ and $\rho$ along $c_2=1.5$
as a function of $c_1$.
Both quantities show hysteresis loops, which are signals of a
first-order phase transition.
In Fig.\ref{fig5}, we present $C$
at $c_2=3.0$. The peak of $C$ at around $c_1\sim 0.48$
develops as $L$ is increased, whereas $E$ shows no 
discontinuity and hysteresis.
Therefore, we conclude that the phase transition at 
$(c_2,c_1)\sim (3.0, 0.48)$ is second order.
In Fig.\ref{fig6}(a), we present the phase diagram of 
the symmetric case for $N=3$, where the order 
of transition between the confinement and Higgs phases changes
from first (smaller $c_2$) to second order (larger $c_2$).  
In Fig.\ref{fig6}(b) we present $C$ along $c_1 = 0.2$, which
shows a smooth nondeveloping peak. $\rho$ decreases smoothly
around this peak.  These results indicate crossovers
at $c_2 \simeq 1.5$.

Then it becomes interesting to consider 
asymmetric cases, e.g., $c_{11}\neq c_{12}=c_{13}$.
This case is closely related to a doped AF magnet. 
$\phi_2$ and $\phi_3$ correspond there to the $CP^1$ spinon
field in the deep easy-plane limit, whereas $\phi_1$ corresponds 
to doped holes. 
This case is also relevant to cosmology because 
the order of Higgs phase transition in the early universe is 
important in the inflational cosmology.
Furthermore, one may naively expect that once a 
phase transition to the Higgs phase occurs at certain 
temperature $T$, 
no further phase transitions take place at lower $T$'s 
even if the gauge field couples with other Higgs bosons.
However, our investigation below 
will show that this is not the case.

\begin{figure}
\includegraphics[width=8cm]{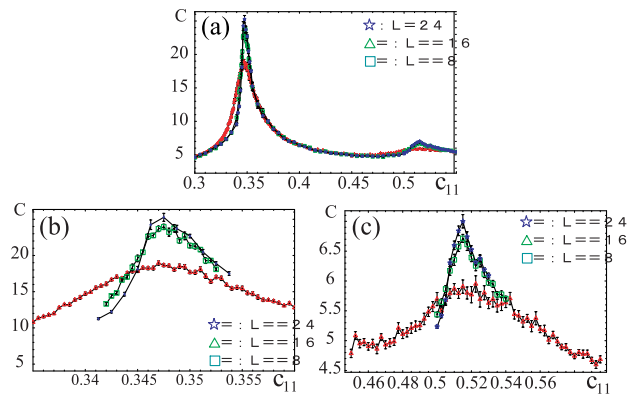}
\caption{\label{fig7}
(a) Specific heat of the $c_1=(1,2,2)$ model ($N$=3) at $c_2=1.0$.
(b,c) Close-up views of $C$ near (b) $c_{11} \sim 0.35$ 
and (c) $c_{11}\sim 0.52$.}
\end{figure}

Let us consider the case $c_{12}=c_{13}=2c_{11}$,
which we call the $c_1=(1,2,2)$ model, and focus on the case
$c_2=1.0$. 
As shown in Fig.\ref{fig7}(a), 
$C$ exhibits two peaks at $c_{11}\sim 0.35$ and $0.52$.
Figs.\ref{fig7}(b),(c) present the detailed behavior 
of $C$ near these peaks, which show 
that the both peaks develop as $L$ is increased.
We conclude that both of these peaks show second-order
transitions. 
This result is interpreted as 
the first-order phase transition 
in the symmetric 
$N=3$ model is decomposed into two second-order transitions
in the $c_1=(1,2,2)$ model.

\begin{figure}
\includegraphics[width=4.5cm]{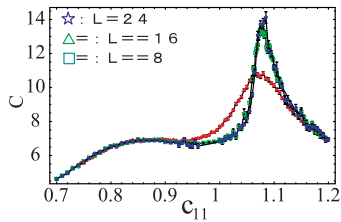}
\vspace{-0.2cm}
\caption{\label{fig8}
Specific heat of the $c_1=(2,1,1)$ model at $c_2=1.0$.}
\end{figure}

Let us turn to the opposite case, $c_{12}=c_{13}=0.5c_{11}$, 
i.e., the $c_1=(2,1,1)$ model at $c_2=1.0$.
One may expect that two second-order phase transitions 
appear as in the previous $c_1=(1,2,2)$ model.
However, the result shown in Fig.\ref{fig8} indicates 
that there exists only one second-order phase transition near 
$c_{11}\sim 1.08$.
The broad and smooth peak near $c_{11}\sim 0.85$ shows no $L$ 
dependence and we conclude that it is a crossover.
This crossover is similar to that 
in the ordinary $N=1$ gauge-Higgs system 
as we shall see by the measurement of $\rho$ below.

The orders of these transitions are understood as follows: 
In the $c_1=(1,2,2)$ model, as we increase $c_{11}$,
the two modes $\phi_{xa} (a=2,3)$ with larger $c_{1a}$ firstly 
become relevant 
and the model is effectively the symmetric $N=2$ model.
The peak in Fig.\ref{fig7}(b) is interpreted as the 
second-order peak of this model. 
For higher $c_{11}$'s, the gauge field is negligible
due to  small fluctuations, and the effective model is the $N=1$  
$XY$ model of  $\phi_{x1}$. It  gives 
 the second-order peak in Fig.\ref{fig7}(c).
Similarly, in the $c_1=(2,1,1)$ model,  $\phi_{x1}$ firstly becomes 
relevant.
The effective model is the   $N=1$ model, which gives 
the broad peak in Fig.\ref{fig8} as 
the crossover\cite{janke}. For higher $c_{11}$'s,
the effective model is the $N=2$ symmetric model of $\phi_{x2},
\phi_{x3}$ and $U_{x\mu}$, giving  the sharp second-order 
peak in Fig.\ref{fig8}.

%

\begin{figure}
\includegraphics[width=8cm]{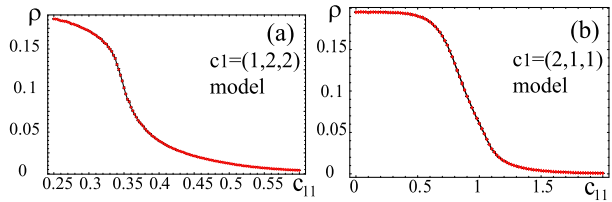}
\vspace{-0.2cm}
\caption{\label{fig9}
Instanton density $\rho$ at $c_2=1.0$ in the (a) $c_1=(1,2,2)$ 
model and (b) $c_1=(2,1,1)$ model.}
\end{figure}


In Fig.\ref{fig9}, we present $\rho$
of the $c_1=(1,2,2)$ and $(2,1,1)$ models
as a function of $c_{11}$.  
$\rho$ of the $c_1=(1,2,2)$ model  
decreases very rapidly at around $c_{11}\sim 0.35$, 
which is the phase transition point in lower $c_{11}$ region.
On the other hand,  at the higher phase
transition point, $c_{11}\sim 0.52$, $\rho$ shows
no significant changes.
This observation indicates that the lower-$c_{11}$ 
phase transition is the confinement-Higgs transition, whereas 
the higher-$c_{11}$ transition is a charge-neutral 
$XY$-type phase transition.

On the other hand, $\rho$ of the $c_1=(2,1,1)$ model 
decreases rapidly at around $c_{11}\sim 0.85$,
where $C$ exhibits a broad peak.
This indicates that the crossover from the dense to 
dilute-instanton regions occurs there just like in 
the $N=1$ case\cite{janke}.
No ``anomalous" behavior of $\rho$ is observed at the critical 
point $c_{11} \sim 1.1$, and therefore the phase transition 
is that of the neutral mode. 

We have also studied the symmetric case for
$N=4,5$ at $c_2=0$. 
Both cases show clear signals of 
first-order transitions at $c_1 \simeq 0.89 (N=4),
0.86 (N=5)$. 
On the other hand, at  $c_2=\infty$,
the gauge dynamics is ``frozen" to $U_{x\mu}=1$
up to gauge transformations, so 
there remain $N$-fold independent $XY$ spin models,
which show a second-order transition at $c_1 \simeq 0.46$.
Thus we expect 
a tricritical point for general $N > 2$ at some {\it finite} $c_2$
separating first-order and second-order transitions. 

Let us summarize the results.
For $N=2$ there is a critical line $\tilde{c}_{1}(c_2)$ of  
second-order transitions in the $c_2-c_1$ plane,
which distinguishes the Higgs phase ($c_1 > \tilde{c}_{1}$)
and the confinement phase ($c_1 < \tilde{c}_{1}$).
This result is consistent with Kragset et al.\cite{kragset}.
For $N=3$ there is a similar transition line, but the region
$0 < c_2 < c_{2c} \simeq 2.25$ is of {\em second-order}
transitions while the region $c_{2c} < c_2$ is
of {\it first-order}  transitions. 
To study the mechanism of generation of these first-order 
transitions, we studied the asymmetric cases and found 
two second-order transitions [in the 
$c_1=(1,2,2)$ model] or one crossover
and one second-order phase transition 
[in the $c_1=(2,1,1)$ model]. 
The former case implies that two simultaneous second-order 
transitions strengthen the order to generate a first-order 
transition.
Chernodub et al.\cite{chernodub} reported
a similar generation of an enhanced first-order transition
in a related 3D Higgs model 
with singly and doubly charged scalar fields. 
We stress that the above  change of the order is dynamical
because (1) It depends on the value of $c_2$,
(2) Related 3D models, the $CP^{N-1}$ and $N$-fold
$CP^1$ gauge models, exhibit always second-order 
transitions (See the last reference of Ref.\cite{gauge}).

We thank Dr.K. Sakakibara for useful discussion.

\end{document}